\documentclass[epsfig,floats,pre,twocolumn,superscriptaddress,
preprintnumbers,floatfix]{revtex4-1}

\usepackage[latin2]{inputenc}
\usepackage{amsmath}
\usepackage{amssymb}
\usepackage{epsfig}
\usepackage{bm}
\usepackage{color}
\usepackage{stackengine}

\begin{document}

\title{Flagellar number governs bacterial spreading and transport efficiency}

\author{Javad Najafi}
\thanks{J.\,N. and M.\,R.\,S. contributed equally to this work.}
\affiliation{Department of Experimental Physics, Saarland 
University, 66123 Saarbr\"ucken, Germany.}
\author{M.\ Reza Shaebani}
\thanks{J.\,N. and M.\,R.\,S. contributed equally to this work.}
\affiliation{Department of Theoretical Physics, Saarland 
University, 66123 Saarbr\"ucken, Germany.}
\author{Thomas John}
\affiliation{Department of Experimental Physics, Saarland 
University, 66123 Saarbr\"ucken, Germany.}
\author{Florian Altegoer}
\affiliation{Department of Chemistry and LOEWE Center for 
Synthetic Microbiology, Philipps University Marburg, 35043 
Marburg, Germany.}
\author{Gert Bange}
\affiliation{Department of Chemistry and LOEWE Center for 
Synthetic Microbiology, Philipps University Marburg, 35043 
Marburg, Germany.}
\author{Christian Wagner}
\email{Corresponding author. Email: c.wagner@mx.uni-saarland.de}
\affiliation{Department of Experimental Physics, Saarland 
University, 66123 Saarbr\"ucken, Germany.}
\affiliation{Physics and Materials Science Research Unit, 
University of Luxembourg, 1511 Luxembourg, Luxembourg.}

\begin{abstract}\noindent
\textbf{Peritrichous bacteria synchronize and bundle their flagella to 
actively swim while disruption of the bundle leads to tumbling. It is 
still not known whether the number of flagella represents an evolutionary 
adaptation towards optimizing bacterial navigation. Here, we study the 
swimming dynamics of differentially flagellated \emph{Bacillus subtilis} 
strains in a quasi-two-dimensional system. We find that decreasing the 
number of flagella {\em N} reduces the average turning angle between 
two successive run phases and enhances the duration and directional 
persistence of the run phase. As a result, having less flagella is 
beneficial for long-distance transport and fast spreading, while having 
a lot of flagella is advantageous for the processes which require 
localization and slow dynamics, such as biofilm formation. We develop 
a two-state random walk model that incorporates spontaneous switchings 
between the states and yields exact analytical expressions for 
transport properties, in remarkable agreement with experiments. The 
results of numerical simulations based on our two-state model suggest 
that the efficiency of searching and exploring the environment is 
optimized at intermediate values of {\em N}. The optimal choice of 
{\em N}, for which the search time is minimized, decreases with 
increasing the size of the environment in which the bacteria swim.}
\end{abstract}
 
\maketitle

\noindent\textcolor{red}{INTRODUCTION}

\noindent
Many bacterial species swim by the rotation of flagella 
\cite{Berg04}. Several flagellation patterns can be distinguished 
according to the flagellar arrangement on the cell body, ranging 
from polar (one flagellum at one pole) to peritrichous (helical 
arrangement on the whole cell body) \cite{Schuhmacher15}. Each 
flagellum is anchored within the cell membrane to a reversible 
rotary motor that switches between clockwise and counter-clockwise 
rotation \cite{Berg73}. The ability of the motor to turn in either 
direction allows {\em polarly-flagellated} bacteria, such as {\em 
Caulobacter crescentus}, to change the swimming direction of the 
cell \cite{Lauga09}. A {\em peritrichously-flagellated} bacterium 
experiences an alternating sequence of active runs and tumbles 
\cite{Berg72}, controlled by the rotational states of each flagellum. 
When all flagella rotate counter-clockwise, they form a bundle and 
synchronize their rotation. As a result, the bacterium moves forward 
smoothly (run phase). However, the bundle is disrupted when some of 
the flagella switch their rotational direction, which facilitates 
changes in the direction of motion in relatively short time intervals 
(tumble phase). The {\em run-and-tumble} dynamics of bacteria allows 
them to change their direction of motion. They can also adjust the 
duration of stay in the run phase in response to environmental 
changes induced, e.g., by temperature, light, or chemical gradients 
\cite{Berg04,Berg72}. Such an ability is known to be highly 
advantageous to enhance search efficiency \cite{Benichou05,
Bartumeus08,Benichou11}, allowing for an optimal navigation towards 
favorable or away from harmful regions \cite{Wadhams04}.  

Despite the advances in understanding of the underlying mechanisms of 
bundle formation and disruption \cite{Chattopahyay06,Darnton07,Mears14,
Turner00,Korobkova06,Terasawa11,Hu13,Hu15} and its relationship to the 
chemotaxis signaling network \cite{Berg04,Wadhams04,Vladimirov10,
Sneddon12}, it remains poorly understood how the number of flagella 
influences the fundamental properties of bacterial propulsion and 
transport. Attempts to understand the role of flagellation have 
mainly focused on {\em Escherichia coli} which has up to eight 
flagella, while the flagellar number of many types of bacteria 
is typically much higher. So far, it has been observed that the 
torque and thus the swimming speed \cite{Darnton07} and the 
fraction of time spent in run or tumble phase \cite{Mears14} 
of peritrichously-flagellated bacteria remain roughly independent 
of the flagellar number $N_{\!f}$. Recent numerical studies also 
predict that the swimming speed is only slightly affected by $N_{\!f}$, 
assuming that just a single bundle is formed \cite{Kanehl14,Reigh12}. 
Here, we provide evidences for the formation of multiple bundles, 
revealing that the former assumption does not hold at least at 
high flagellar number.

A key question from an evolutionary point of view is based on 
which criteria a bacterium chooses to have a certain number of 
flagella. In the present study, we investigate bacterial motility 
over a wide range of the flagellar number and clarify how the 
run-and-tumble dynamics is influenced by the choice of $N_{\!f}$. 
We study \emph{Bacillus subtilis}, a rod-shaped bacterium commonly 
found, e.g., in soil and the gastrointestinal tract of humans. The 
number of flagella is regulated by the master regulator swrA in 
\emph{Bacillus subtilis} \cite{Kearns05}. Deletion of the gene 
leads to a reduction from $26$ to $10$ flagella, while cells 
overexpressing swrA exhibit up to $40$ flagella. Therefore, we 
employed a wild-type \emph{Bacillus subtilis} (NCIB3610 strain) 
with $26$ flagella, a $\Delta\text{swrA}$ strain deficient for 
swrA ($9$ flagella), and a one which carries swrA under the 
control of an IPTG-inducible promoter ($41$ flagella) (Fig.\,1A). 
While the speeds in both run and tumble phases and the mean 
duration of tumbling periods show no systematic dependence 
on $N_{\!f}$, the duration and directional persistence of 
the run phase and the average turning angle between two 
successive run phases vary monotonically with $N_{\!f}$. 
Smaller values of $N_{\!f}$ are more favorable for 
long-distance transport and fast spreading because of 
more persistent trajectories and longer excursion times 
in the run phase. On the other hand, increasing $N_{\!f}$ 
facilitates tumbling, leading to slow dynamics and localization 
which is beneficial e.g.\ for the formation of biofilms. We 
develop a coarse-grained analytical approach to study the 
run-and-tumble dynamics of bacteria, which enables us to 
identify the contributions of the influential factors and 
map out the phase diagrams in the space of the relevant 
parameters of run-and-tumble statistics. Our simulation 
results suggest that by adopting intermediate values of 
$N_{\!f}$ (as e.g.\ in the wild-type strain), the search 
efficiency and the ability to explore the environment 
can be optimized.\\ 

\noindent\textcolor{red}{RESULTS}

\noindent
We report and compare the dynamics of three strains of 
\emph{Bacillus subtilis} with different flagellar number 
$N_{\!f}$ obtained by genetic manipulation (see 
\emph{Materials and Methods} section for details). The 
strains $\Delta\text{swrA}$, $\text{wt}$ (wild-type 
NCIB3610) and $\text{swrA}$ have $9{\pm}2$, $26{\pm}6$ 
and $41{\pm}6$ flagella, respectively \cite{Guttenplan13} 
(Fig.\,1A). In contrast to the studies on tumbling of 
trapped cells \cite{Mears14}, here we track the motion 
of freely swimming cells. A few typical trajectories 
of each strain are shown in Fig.\,1B, which consist 
of successive active run and tumble phases.\\ 

\begin{figure}[t]
\centerline{\includegraphics[width=0.45\textwidth]{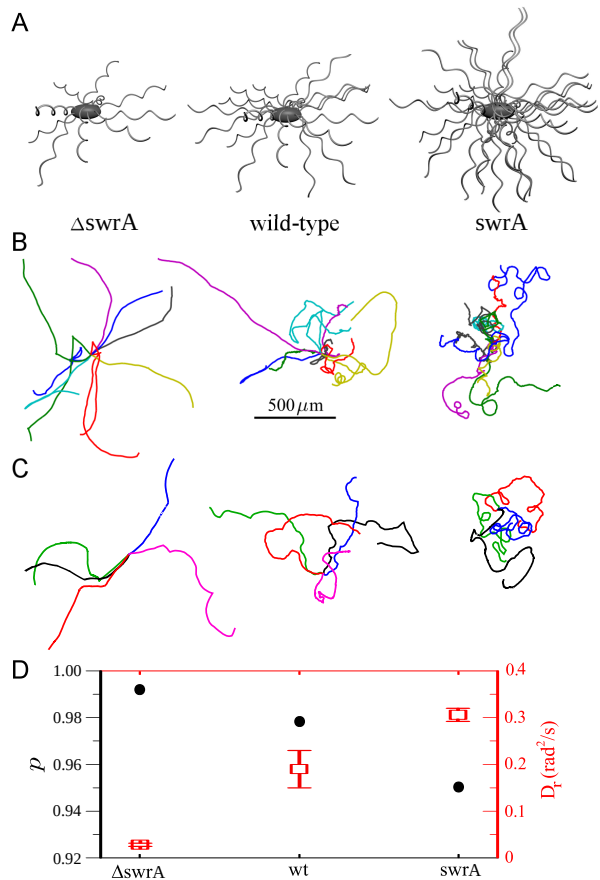}}
\caption{\textbf{Bacterial motility patterns.} (A) Illustration 
of three strains of \emph{Bacillus subtilis} with different number 
of flagella. (B),(C) Planar projections of quasi two-dimensional 
experimental (B) and simulation (C) paths. A few trajectories are 
randomly chosen and translated so that the first point is located 
at the origin. The simulation parameter values are extracted from 
experimental data. (D) Directional persistency of the run phase $p$ 
(circles), and asymptotic rotational-diffusion coefficient $D_r$ 
(squares) for different strains.}
\label{Fig:1}
\end{figure}

\noindent\textbf{Distinct motility patterns} 

\noindent
Strikingly, we find that the curvatures of the trajectories 
strongly depend on the flagellar number. The paths are clearly 
less curved with decreasing $N_{\!f}$, which may originate either 
from the change in their tumbling statistics or from having a 
larger persistence length in the run phase. While the influence 
of $N_{\!f}$ on the run-and-tumble statistics will be discussed 
in the next section, here we focus on the latter possibility, i.e.\ 
the variations of the persistency in the run phase with changing 
$N_{\!f}$. To characterize the running persistency, we note that 
each run trajectory comprises a set of recorded positions of the 
bacterium. Given these data, one can extract the local direction 
of motion $\alpha$ and, thus, the directional change $\theta$ along 
the trajectory, resulting in the distribution of directional changes 
$P(\theta)$ \cite{Burov13}. Then, the persistency $p$ of the run 
phase can be obtained as the Fourier transform of $P(\theta)$, 
i.e.\ $p\!=\!\int_{-\pi}^{\pi}\text{d}\theta\;e^{\text{i}\theta}
P(\theta){=}\langle\cos\theta\rangle$ \cite{Shaebani14,Sadjadi15}. 
Alternatively, the curvature of the trajectories can be characterized 
by calculating the rotational mean square displacement $\text{RMSD}
(\tau){=}\big\langle\big(\alpha(t{+}\tau){-}\alpha(t)\big)^2\big
\rangle{=}2\,D_r\,\tau$, from which one can extract the asymptotic 
rotational-diffusion coefficient $D_r$. The results shown in 
Fig.\,1D reveal that the directional persistency in the run phase 
remarkably depends on the flagellar number; the run paths are 
less curved for smaller $N_{\!f}$.\\ 

\begin{figure*}
\centerline{\includegraphics[width=0.97\textwidth]{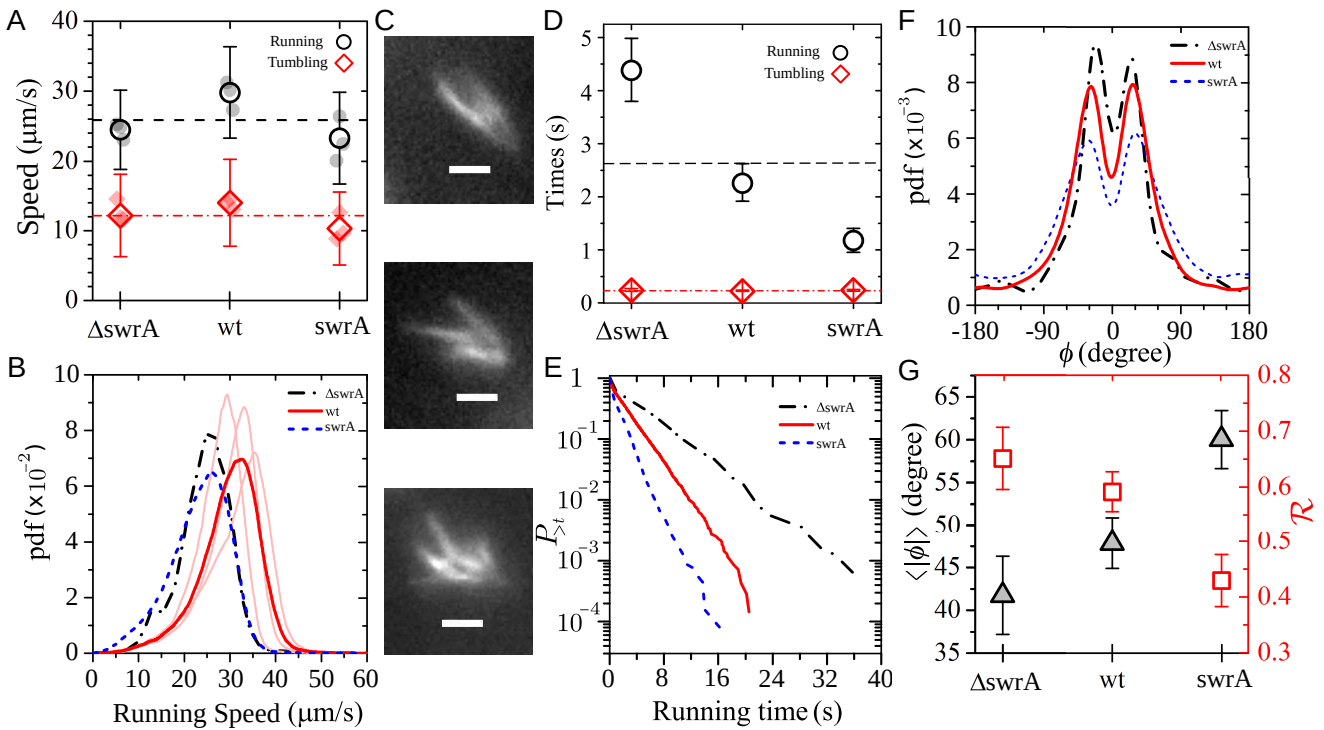}}
\caption{\textbf{Run-and-tumble statistics.} (A) The mean run 
and tumble speeds for different strains (open symbols). The 
full symbols indicate the results for different cultures of each 
strain. The horizontal lines show the mean values over all strains. 
(B) The probability distribution of the run speed. The thin solid 
curves (bright red) indicate the distributions for different 
cultures of $\text{wt}$ strain. (C) Fluorescence images of the 
bundles of a wild-type strain. Scale bar, $3.5\,\mu\text{m}$. 
(D) The average duration of run (circles) and tumble (diamonds) 
periods. The horizontal lines indicate the average values over 
all strains. (E) The probability distribution of observing a 
running time longer than a given duration $t$. (F) The 
turning-angle probability distribution $R(\phi)$. (G) The 
average turning angle $\langle|\phi|\rangle$ (triangles) and 
$\mathcal{R}{=}\langle\cos\phi\rangle$ (squares) for different 
strains.}
\label{Fig:2}
\end{figure*}

\noindent\textbf{Run-and-tumble statistics}

\noindent
Next, we turn to studying the statistics of run and tumble events. 
A detailed description of the tumbling detection procedure can be 
found in \emph{Materials and Methods} section. According to the results 
shown in Fig.\,2A, the tumbling speed is smaller (but non-negligible) 
compared to the running speed \cite{Turner16}. Moreover, the speeds 
in both run and tumble phases show no systematic dependence on 
$N_{\!f}$. A closer look at the probability distributions of the 
instantaneous speeds (see e.g.\ the running speed data in Fig.\,2B) 
reveals that even the distributions of a given strain slightly 
differ from one culture to another due to biological heterogeneity. 
Thus, there is no significant difference in the speeds of different 
strains. By approximating the bundle as a single helix \cite{Lauga09,
Kanehl14,Rodenborn13}, simple models based on resistive-force theory 
predict a weak (logarithmic) growth of running speed with $N_{\!f}$, 
though experiments have shown that hydrodynamic dissipation along 
the cell body considerably weakens the effect \cite{Darnton07}. 
Additionally, the assumption of a single bundle does not hold in 
general, according to our observations of fluorescently-stained 
flagella of swimming bacteria as well as other reports in the 
literature \cite{Valeriani11,Hyon12}; even multiple bundles may 
form during the active running phase (several examples of a 
$\text{wt}$ strain with multiple bundles are shown in Fig.\,2C). 
The diversity of the locations and interactions between the 
bundles makes the prediction of the propulsive force and 
swimming speed highly complex.

To understand whether and to what extent the flagellar number 
influences the switching frequencies between run and tumble 
phases, we plot the average excursion times in Fig.\,2D. 
It turns out that the mean tumbling time (${\sim}0.2\,\text{s}$) 
and thus the ability of restoring the run phase are not influenced 
by $N_{\!f}$. Given the mean tumbling time and speed, one 
finds that the displacement in the tumbling phase is of the 
order of the cell length (i.e.\ a few micrometers). However, 
such a swimmer with the running speed of ${\sim}30\,\mu
\text{m}{/}\text{s}$ in an aqueous medium is expected to 
stop after only ${\sim}0.6\,\mu\text{s}$ and traveling 
approximately $0.1\,A^{\circ}$ if the pushing force suddenly 
stops \cite{purcell77}. This suggests that the bundles are 
not fully disrupted in the tumbling phase, thus, the 
bacterium is partially propelled. The mean running times 
remarkably depend on the flagellar number. Increasing 
$N_{\!f}$ enhances the probability of switching from 
running to tumbling. This supports the \emph{veto} model 
for bundle disruption \cite{Turner00,Sneddon12}, which 
proposes that clockwise rotation of a single or a few 
flagella (less than $\frac{N_{\!f}}{2}$) is sufficient to 
disrupt the bundle. Figure\,2E shows that the tail of the 
probability distribution of running times is nearly exponential, 
with a $N_{\!f}$-dependent slope (data censoring for the 
first and last incomplete periods sharpens the trend even 
more). We observe similar exponential decays for the 
tumbling time distributions (not shown), verifying that 
the switching events between the two states of motility 
happen spontaneously and can be described by Markovian 
processes.  

Another characteristic of the run-and-tumble dynamics that 
is affected by the number of flagella is the turning angle 
between successive run phases. We define the turning angle 
$\phi$ between two successive runs as the change in the 
direction of motion from the end of one run to the beginning 
of the next run (the direction of motion was obtained by a 
linear fit to four data points for each run phase). The 
turning-angle distribution $R(\phi)$ obeys left-right 
symmetry and develops a peak for all strains. However, the 
peak and, more clearly, the mean of $R(\phi)$ shift towards 
larger angles with increasing $N_{\!f}$, as can be seen in 
Figs.\,2F and 2G. Similar to the curvature of run 
trajectories, we can use the Fourier transform of 
$R(\phi)$, i.e.\ $\mathcal{R}\!=\!\int_{-\pi}^{\pi}
\text{d}\phi\;e^{\text{i}\phi}R(\phi){=}\langle\cos
\phi\rangle$, as a measure of the directional change 
between consecutive run phases. $\mathcal{R}$ ranges 
between $-1$ and $1$, with $1$, $0$, and $-1$ denoting 
$0^{\circ}$, $90^{\circ}$, and $180^{\circ}$ turning, 
respectively. Indeed, $\mathcal{R}$ is correlated with 
the tumbling time and speed \cite{saragosti12} as well 
as the strength of the torque exerted on the cell body 
during the re-formation of the bundle \cite{Darnton07}. 
Since the speed and excursion time in the tumbling phase 
are not considerably affected by $N_{\!f}$, we attribute 
the reduction of $\mathcal{R}$ with increasing $N_{\!f}$ 
to experiencing a larger torque during bundle re-formation 
at higher flagellar numbers \cite{Darnton07}. 

We also checked that the direction of motion does not vary 
significantly when switching from running to tumbling. 
This fact and the relatively high speed in the tumbling 
phase once again indicate that the bundles are not fully 
disrupted in the tumbling phase. As a result, one expects 
that the tumbling is not a purely random motion and should 
exhibit directional persistency. To verify this idea we 
assign a dimensionless number $\tilde\ell_{_\text{tumble}}$ 
to each tumbling segment of the trajectory, obtained by 
dividing the end-to-end distance by the total length of 
the trajectory segment. The probability distribution 
$P(\tilde\ell_{_\text{tumble}})$ shown in Fig.\,3A 
develops a sharp peak at $\tilde\ell_{_\text{tumble}}
{\sim}0.9$, indicating that most of the tumbling 
trajectories are nearly straight during the relatively 
short periods of tumbling.

\begin{figure}[t]
\centerline{\includegraphics[width=0.48\textwidth]{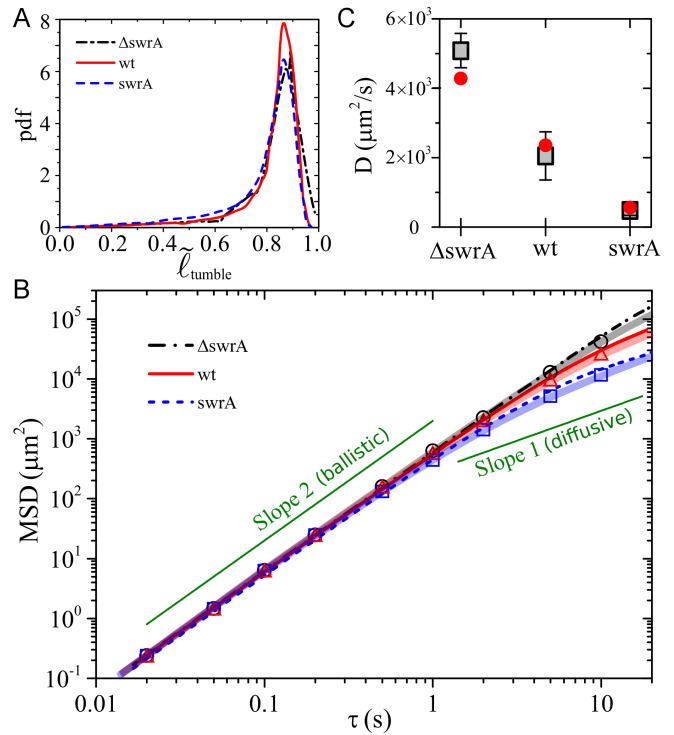}}
\caption{\textbf{Spreading and transport properties.} (A) 
Probability distribution of the dimensionless quantity 
$\tilde\ell_{_\text{tumble}}$, reflecting the bending of tumbling 
trajectories. (B) Evolution of the mean square displacement for 
different strains, obtained from experiments (thin dark-color lines), 
simulations (symbols), and the model via Eq.\,\ref{Eq:MSDz} (thick 
bright-color lines). (C) The asymptotic diffusion coefficient $D$ 
for different strains. The analytical predictions (circles) are 
compared to the experimental results (squares).}
\label{Fig:3}
\end{figure}

Finally, we measure the mean square displacement $\text{MSD}
(\tau){=}\big\langle\big(r(t{+}\tau){-}r(t)\big)^2\big\rangle$ 
for different strains. According to our findings, the strains 
with lower flagellar number have less-curved run trajectories, 
switch less frequently from run to tumble phase, and experience 
a smaller turning angle between successive run phases. Consequently, 
the crossover from a persistent motion to the asymptotic diffusive 
dynamics is expected to occur at longer times for smaller values of 
$N_{\!f}$, which is confirmed by the experimental results shown 
in Fig.\,3B. The asymptotic diffusion coefficient $D$ is also 
a decreasing function of $N_{\!f}$ (see Fig.\,3C). These results 
indicate that for increasing the efficiency of long-distance 
transport, it is more beneficial to have less flagella. The 
strains with more flagella spend a larger fraction of time 
in tumbling state and have a shorter persistence length when 
swimming. We note that the trends (as a function of $N_{\!f}$) 
discussed in this section are robust against variations of 
tumbling detection thresholds. \\

\noindent\textbf{Model}

\noindent
Based on our findings, we develop a stochastic coarse-grained 
model for a random walk with spontaneous switchings between 
two states of motility. We first validate the approach by 
comparing the analytical predictions with experimental 
results. Then the model enables us to better understand 
the influence of the flagellar number on bacterial motility, 
by numerically exploring the phase space of the run-and-tumble 
statistical parameters and identifying the isolated role 
of each influential factor. Moreover, our simulations based 
on the two-state model evidence for a nonmonotonic dependence 
of the mean first-passage time on $N_{\!f}$.

\begin{figure*}
\centerline{\includegraphics[width=0.94\textwidth]{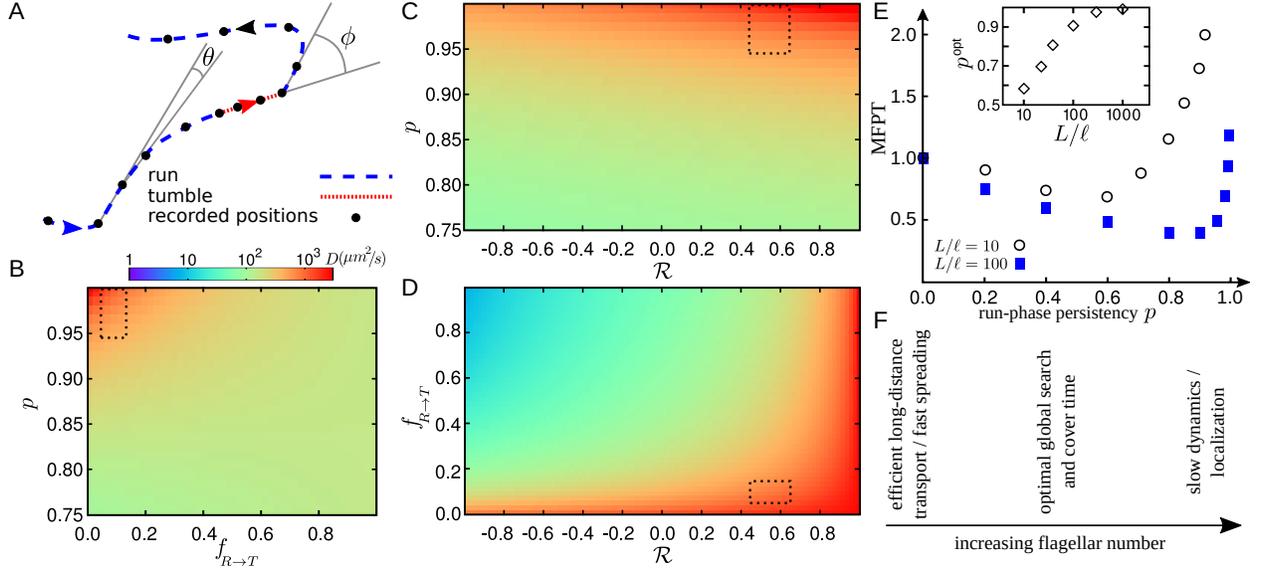}}
\caption{\textbf{Phase diagram of the influential parameters.} 
(A) A typical sample trajectory of \emph{Bacillus subtilis} 
with run-and-tumble dynamics. (B)-(D) Three cross-sections 
corresponding to (B) $\mathcal{R}{=}\langle\cos\phi\rangle{=}0.56$, 
(C) $f_{_{_{\text{R}{\!\rightarrow\!}\text{T}}}}{=}0.05$, 
and (D) $p{=}\langle\cos\theta\rangle{=}0.97$ of the three 
dimensional phase diagram in the ($\mathcal{R}$, $p$, 
$f_{_{_{\text{R}{\!\rightarrow\!}\text{T}}}}$) space. The 
color intensity reflects the magnitude of the asymptotic 
diffusion coefficient $D$. The marked regions indicate the 
accessible range of parameters in our experiments with 
\emph{Bacillus subtilis}. (E) Mean first-passage time MFPT, 
scaled by MFPT at $p{=}0$ versus the directional persistency 
$p$ of the run phase. (inset) Optimal persistency $p^{\text{opt}}$ 
vs the effective system size $L{/}\ell$. (F) Schematic picture 
depicting how different aspects of transport efficiency vary 
with $N_{\!f}$.}
\label{Fig:4}
\end{figure*}

Stochastic two-state models have been widely employed to 
describe altering phases of motion of e.g.\ swimmers and 
cytoskeletal motor proteins, or the locomotive patterns in 
other systems \cite{Theves13,Taktikos13,Hafner16,Pinkoviezky13,
Shaebani17,Kong15,Watari10}. Aiming at obtaining analytical 
insight by including only the most prominent characteristics 
of the bacterial motility pattern into the model, we introduce 
a random walk approach in which the walker experiences two 
states of motility: (i) running with the mean speed 
$v_{_\text{R}}$ and persistency $p$, and (ii) tumbling 
with the mean speed $v_{_\text{T}}$. The switching 
probabilities from run to tumble state and vice versa are 
supposed to be asymmetric and denoted by 
$f_{_{_{\text{R}{\!\rightarrow\!}\text{T}}}}$ and $f_{_{_{\text{T}{
\!\rightarrow\!}\text{R}}}}$, respectively. Assuming constant 
switching probabilities results in exponential distributions 
for the residence times in each state, which is in agreement 
with our experimental observations. In the run state, a 
persistent random walk in the continuous space is considered 
which is characterized by its speed $v_{_\text{R}}$ and 
the distribution $P(\theta)$ of directional changes along 
the run trajectory. Thus, the directional persistence $p{=}
\int_{-\pi}^{\pi}\text{d}\theta\;e^{\text{i}\theta}P(\theta)
{=}\langle\cos\theta\rangle$ quantifies the curvature of 
the run trajectories. Based on the results shown in Fig.\,3A, 
the tumbling trajectories are approximated by straight lines 
along the last direction of motion in the run phase (see 
Fig.\,4A). The last key property is the turning 
angle $\phi$ between two successive run phases. Using the 
turning-angle distribution $R(\phi)$, we calculate 
$\mathcal{R}{=}\int_{-\pi}^{\pi}\text{d}\phi\;e^{\text{i}
\phi}R(\phi){=}\langle\cos\phi\rangle$ to quantify the 
directional change between consecutive run phases. Denoting 
the time spacing between consecutive frames by $\Delta t$, 
we describe the process in discrete time by introducing 
the probability densities $P_{t}^{R}(x,y|\gamma)$ and 
$P_{t}^{T}(x,y|\gamma)$ to find the walker at position 
$(x,y)$ along the direction $\gamma$ at time $t$ in the 
run and tumble states, respectively. The dynamical evolution 
is defined by the following set of master equations 
\begin{equation}
\begin{array}{ll}
P_{\!_{t{+}\Delta t}}^{R}(x,y|\gamma) = \vspace{1mm} \\
\,(1{-}f_{_{_{\text{R}{\!\rightarrow\!}\text{T}}}}) 
\!\!\! \displaystyle\int_{\!{-}\pi}^{\pi} \!\!\!\!\!\!\! 
d\beta P(\gamma{-}\beta) 
P_{t}^{R}(x{-}v_{_\text{R}}\Delta t\cos\gamma,y{-}
v_{_\text{R}}\Delta t\sin\gamma|\beta) \vspace{1mm} \\
\,+ f_{_{_{\text{T}{\!\rightarrow\!}\text{R}}}} 
\! \displaystyle\int_{\!{-}\pi}^{\pi} \!\!\!\!\! 
d\beta \, R(\gamma{-}\beta) 
\, P_{t}^{T}(x{-}v_{_\text{R}}\Delta t\cos\gamma,y{-}
v_{_\text{R}}\Delta t\sin\gamma|\beta), \vspace{3mm} \\
P_{\!_{t{+}\Delta t}}^{T}(x,y|\gamma) = \vspace{1mm} \\
\,\,\,\,\,(1{-}f_{_{_{\text{T}{\!\rightarrow\!}\text{R}}}}) 
\, P_{t}^{T}(x{-}v_{_\text{T}}\Delta t\cos\gamma,y{-}
v_{_\text{T}}\Delta t\sin\gamma|\gamma) \vspace{1mm} \\
\,\,\,\,\,+ f_{_{_{\text{R}{\!\rightarrow\!}\text{T}}}} 
\, P_{t}^{R}(x{-}v_{_\text{T}}\Delta t\cos\gamma,y{-}
v_{_\text{T}}\Delta t\sin\gamma|\gamma).
\end{array}
\label{Eq:MasterEqs}
\end{equation}
By the two terms on the right-hand side of each equation, we 
consider the possibilities of being at each of the two states 
in the previous step. On the right-hand side of the second 
equation, the possibilities of moving along a straight line 
or switching from running to tumbling are considered. For 
simplicity we assume that the direction of motion does not 
change when switching from run to tumble state. By means of 
a Fourier-z-transform technique \cite{Sadjadi08,Sadjadi11}, 
we obtain exact analytical expressions for temporal evolution 
of arbitrary moments of displacement (see \emph{Materials 
and Methods} section for the details of the method and the 
explicit lengthy expression of MSD). 

\begin{table}[b]
\centering
\begin{tabular}{lr}
Table 1: Parameter values extracted from experiments.\\
\end{tabular}
\begin{tabular}{l*{5}{c}c} 
strain&$v_{_\text{R}}\!(\frac{\mu\text{m}}{\text{s}})$&$v_{_\text{T}}
\!(\frac{\mu\text{m}}{\text{s}})$&$\langle t_{_\text{R}} \rangle(\text{s})$&
$\langle t_{_\text{T}} \rangle(\text{s})$&$p$&$\mathcal{R}$\vspace{1mm} \\
\hline
$\Delta\text{swrA}$ & 24.5 & 12.1 & 4.39 & 0.244 & 0.99 & 0.65 \\
$\text{wt}$         & 29.8 & 14.0 & 2.27 & 0.224 & 0.98 & 0.59 \\
$\text{swrA}$       & 23.3 & 10.3 & 1.18 & 0.240 & 0.95 & 0.43 \\
\end{tabular}
\end{table}

The parameter values extracted from experiments are given in 
Table\,1 (for the calculation of $p$ we chose $\Delta\,t{=}
\frac16\,\text{s}$). Using these data as the input of the 
model, the time evolution of the mean square displacement 
can be predicted via Eq.\,\ref{Eq:MSDz} (plotted in Fig.\,3B 
for different strains). We also performed extensive Monte Carlo 
simulations of a stochastic process with the same parameter 
values. The results agree perfectly with the analytical 
predictions. The agreement with experiments is also 
strikingly good; Despite all the simplifications, the model 
recovers the MSDs of all strains over the whole range of time. 
By introducing $A{\equiv}2\,\frac{f_{_{_{\text{R}{\!
\rightarrow\!}\text{T}}}}}{f_{_{_{\text{T}{\!\rightarrow\!}
\text{R}}}}}\,v^2_{_\text{T}}{-}f_{_{_{\text{T}{\!\rightarrow
\!}\text{R}}}}v^2_{_\text{R}}$ and $B{\equiv}2 \,f^2_{_{_{
\text{R}{\!\rightarrow\!}\text{T}}}} \mathcal{R} \,v^2_{_
\text{T}}{+}2 \, f^2_{_{_{\text{T}{\!\rightarrow\!}
\text{R}}}}v^2_{_\text{R}}$, we obtain the following 
expression for the asymptotic diffusion coefficient 
\begin{equation}
D = \frac14 \frac{\Delta\,t}{f_{_{_{\text{T}{\!\rightarrow\!}
\text{R}}}}{+}f_{_{_{\text{R}{\!\rightarrow\!}\text{T}}}}} 
\Bigg[A+\frac{B{+}2\,f_{_{_{\text{T}{\!
\rightarrow\!}\text{R}}}}f_{_{_{\text{R}{\!\rightarrow\!}
\text{T}}}}(1{+}\mathcal{R})v_{_\text{R}}v_{_\text{T}}}{
f_{_{_{\text{T}{\!\rightarrow\!}\text{R}}}}\Big[1{-}p{-}
f_{_{_{\text{R}{\!\rightarrow\!}\text{T}}}}(\mathcal{R}
{-}p)\Big]}\Bigg].
\label{Eq:Dasymp}
\end{equation}
The measured diffusion coefficients for different strains 
are in satisfactory agreement with the analytical predictions, 
as shown in Fig.\,3C. Equation (\ref{Eq:Dasymp}) reduces to 
the asymptotic diffusion coefficient of a single-state 
persistent random walker for $f_{_{_{\text{R}{\!\rightarrow
\!}\text{T}}}}{=}0$ and $f_{_{_{\text{T}{\!\rightarrow\!}
\text{R}}}}{=}1$ \cite{Tierno16}. 

While it is extremely difficult to systematically vary the 
flagellar number in experiments, we can numerically explore 
the full phase space of the influential parameters. Among the 
key elements in determining the transport properties of the 
\emph{Bacillus subtilis}, the speeds $v_{_\text{R}}$ and 
$v_{_\text{T}}$, and the switching probability $f_{_{_{
\text{T}{\!\rightarrow\!}\text{R}}}}$ do not vary significantly 
with $N_{\!f}$. Thus, we fix them at their average experimental 
values, which allows us to reduce the degrees of freedom to 
three sensitive parameters, i.e.\ $\mathcal{R}$, the directional 
persistency $p$, and the switching probability $f_{_{_{
\text{R}{\!\rightarrow\!}\text{T}}}}$. Figures 4B-4D show 2D 
profiles of $D$ in the ($\mathcal{R}$, $p$, $f_{_{_{\text{R}{
\!\rightarrow\!}\text{T}}}}$) phase space, revealing that $D$ 
varies by several orders of magnitude by varying these key 
parameters. However, limiting the parameter values to the 
accessible range in experiments with \emph{Bacillus subtilis} 
reveals that here $D$ and the long-distance transport are 
mainly affected by the variation of the run-phase persistency 
$p$. For example, by fixing the two other degrees of freedom 
at their average experimental values and calculating the 
variations of the asymptotic diffusion coefficient when the 
third parameter varies within the accessible range in our 
experiments, we get $\frac{D_\text{max}}{D_\text{min}}{=}1.2, 
1.8,$ and $3.6$ for varying $\mathcal{R}$, $f_{_{_{\text{R}{
\!\rightarrow\!}\text{T}}}}$, and $p$, respectively. More 
generally, by changing the flagellar number of other types 
of bacteria one may deal with other regions of the 6-fold 
phase space. 

Our analytical approach allows us to extract further information 
about the transport properties, such as the crossover time $t_c$ 
to asymptotic diffusive regime as a function of the key parameters. 
The crossover time can be estimated by balancing the linear terms 
in time and the nonlinear contribution in the MSD equation. We find 
that $t_c$ also varies by several orders of magnitude in the 
($\mathcal{R}$, $p$, $f_{_{_{\text{R}{\!\rightarrow\!}\text{T}}}}$) 
phase space (not shown). Note that due to the interplay between the 
key parameters, multiple transitions between different types of 
anomalous diffusive dynamics occur on varying time scales in general 
\cite{Hafner16}.

Finally, we study the mean first-passage time (MFPT) of a two-state 
persistent random walk in confinement by means of extensive Monte 
Carlo simulations. The MFPT is defined as the mean time taken by the 
random walker to reach a particular position in the system for the 
first time. For simplicity, random walk on a cubic lattice with 
periodic boundary conditions is considered, and all run-and-tumble 
parameters except $p$ (i.e.\ the most sensitive factor upon varying 
$N_{\!f}$) are fixed at their mean experimental values. The walker 
is supposed to search for hidden targets during its both states of 
motility. The results reveal that the MFPT is a nonmonotonic function 
of $p$ and admits a minimum (Fig.\,\ref{Fig:4}E). From the monotonic 
dependence of $p$ on $N_{\!f}$ (Fig.\,\ref{Fig:1}D), one concludes 
that the search efficiency is optimized at intermediate values of 
$N_{\!f}$. Denoting the size of the confinement with $L$, and the 
reaction range of the searcher with $\ell$, the efficient persistency 
grows with increasing $L{/}\ell$ (inset of Fig.\,\ref{Fig:4}E); thus, 
the optimal number of flagella $N_{\!f}^{\text{opt}}$ shifts towards 
smaller values with increasing the effective system size.\\ 

\noindent\textcolor{red}{DISCUSSION}

\noindent
Our investigation of the dynamics of \emph{Bacillus subtilis} 
revealed that the strains with lower flagellar number $N_{\!f}$ 
have higher persistency and longer excursion time in the run 
phase and, additionally, their abrupt directional change when 
switching back to the run phase is smaller; Consequently, 
having too few flagella enhances the efficiency of long-distance 
transport (reflected e.g.\ in their higher translational diffusion 
coefficient), and causes fast spreading. Increasing the number 
of flagella considerably increases the probability of switching 
from running to tumbling as well as the curvature of the 
trajectories in the run phase; Moreover, switching from 
tumbling to running is accompanied by a larger directional 
change when the bacterium has more flagella. As a result, 
the overall orientation changes more frequently if there 
are too many flagella. This leads to slow dynamics and can 
be beneficial for the processes which require localization, 
such as the formation of biofilms.

However, the flagellar number of the wild-type \emph{Bacillus 
subtilis} falls in the middle of the range, where the search 
efficiency could be optimized according to our simulation 
results (see Fig.\,4F). Intermittent random walks are known 
to be beneficial for search efficiency in confined geometries 
\cite{Benichou06,Chabaud15,Rupprecht16}. For example, when 
slow diffusion periods are interrupted by fast relocating 
ballistic flights, it has been shown that there exists an 
optimal ratio between the excursion times in the two states 
which leads to a global minimum of the search time 
\cite{Benichou06}. The optimal choice varies with the system 
size and the reaction range of the searcher. In addition, it 
has been recently shown for a single-state persistent random 
walk, that the mean first-passage time to find a target 
admits a minimum as a function of the persistency \cite{Tejedor12}. 
In a general two-state random walk with tumbling and persistent 
running phases (such as the motion of \emph{Bacillus subtilis}), 
one expects that further complexity to determine the optimal 
search time arises due to the interplay between the directional 
persistency $p$ of the run phase, the switching probabilities 
$f_{_{_{\text{R}{\!\rightarrow\!}\text{T}}}}$ and $f_{_{_{
\text{T}{\!\rightarrow\!}\text{R}}}}$ between run and tumble 
(which determine the ratio between the excursion times in the 
two states), the size of the confinement, and the characteristics 
of the bacterial chemotaxis system (esp.\ the reaction range). 
However, from our experimental findings we expect that changing 
the flagellar number of \emph{Bacillus subtilis} influences the 
dynamics and, thus, the search time mainly via affecting the 
run-phase persistency $p$. If the number of flagella of wild-type 
\emph{Bacillus subtilis} has been solely evolved towards optimizing 
the search efficiency, then our simulation results suggest that 
it should be most appropriate to live in environments with a 
typical size of the order of a few hundreds larger than the 
reaction radius of the bacterium (e.g.\ an environment size 
of less than a millimeter for a reaction radius comparable 
to the typical size of \emph{Bacillus subtilis} i.e.\ a few 
microns). Our results thus provide a new insight into the nature 
of bacterial motility, which goes beyond the mechanochemical 
description of chemotaxis and mechanisms of flagellar bundle 
formation and disruption.\\

\noindent\textcolor{red}{MATERIALS AND METHODS}\\
\noindent\textbf{Genetic manipulation of bacteria}\\
\noindent Allelic replacement was performed using the pMAD-system 
according to [43]. All strains originated from the recently described 
NCIB3610 \emph{Bacillus subtilis} strain harbouring a point mutation 
in the $\text\it{comI}$ gene [44]. Briefly, the $\text\it{hag}$ gene 
encoding flagellin was amplified including up- and downstream flanking 
regions and cloned into the pMAD plasmid. The T209 to C mutation was 
obtained by quick-change mutagenesis. NCIB3610 $\text\it{comI}\!_{_
\text{Q12I}}$ was transformed and positive clones selected according 
to [43]. Exchange of the native $\text\it{hag}$ gene to the $\text
\it{hag}_{_{\text{T209C}}}$ was confirmed by sequencing and light 
microscopy. The strain is referred to as wild-type NCIB3610 in terms 
of the flagellar number in the text. We are grateful to Daniel Kearns 
for strain DK1693 ($\Delta\text{swrA}$, $\text\it{amyE}{::}\text{P}
\!\!_{_\text{hyspank}}\!\!{-}\text{swrA}$; $\text\it{lacA}{::}\text{P}
\!_{_\text\it{hag}}\!{-}\text\it{hag}_{_{\text{T209C}}}$ mls). This 
strain was used as both $\Delta\text{swrA}$ and $\text{swrA}\!^{++}$ 
when induced with IPTG.\\   

\noindent\textbf{Sample preparation}\\
\noindent The temperature was set to 37$^{\circ}$C for all 
steps of the experiment. First, 20 ml of frozen stock \emph{Bacillus 
subtilis} was streaked onto a LB-agar plate. The plate was incubated 
for $\sim$16 hours and then a few colonies of bacteria from the plate 
were stirred in LB and grown over night. The cultures were diluted to 
$OD_{600}{\sim}$0.1 next morning and grown for two more hours to reach 
the early exponential phase within the optical density range of $\sim$0.5 
and 0.8. After dilution, $\text{swrA}$ strain was induced by 1 mM 
IPTG solution to synthesize more flagella. The optical density of 
the cells in the exponential phase was first adjusted to 0.5 and 
then further diluted ($\sim$15 folds) in a fresh LB that had been 
previously purified by 0.4 $\mu$m syringe filter. Because of strong 
sticking of $\Delta\text{swrA}$ strain to the surface, $0.005\%$ 
PVP-40 (polyvinylpyrrolidones) was added before the experiments [45]. 
For fluorescent labeling of cells, the dye was prepared by solving 
1 mg Alexa Fluor$\textregistered$568 C5-maleimide in 200 $\mu$l DMSO 
(Dimethyl sulfoxide). 1 ml of the cell culture in $OD_{600}{\sim}$1 
was centrifuged at 8000g for one minute and gently washed three 
times in PBS pH 7.4(1X). The pellet was resuspended in 200 $\mu$l 
PBS together with 5 $\mu$l of the dye solution. The suspension was 
mixed and incubated in the dark at room temperature for 20 $\text{min}$ 
and washed three times again in PBS buffer and re-energized by 
half hour outgrowth in LB to observe motile cells.

Fluorescence microscopy samples were prepared by adding 30 $\mu$l 
of labeled cell suspension into a FluoroDish FD35-100 and covering 
it by circular cover glass (VWR, diameter 22 mm, thickness NO.\ 
1.5). A Nikon Eclipse Ti microscope together with a Nikon N Plan 
Apo $\lambda$ 60x, N.A.\ 1.4 oil immersion objective were used 
for fluorescence microscopy. The dye was excited by a mercury 
lamp and images were acquired using Hamamatsu ORCA-Flash 4.0 
camera with 25 ms exposure time and $2{\times}2$ binning. Tracking 
chamber was a superposition of a cover slip (VWR, $20{\times}20$ 
mm, thickness NO.\ 1) on a microscope slide (Carl Roth GmbH, 
Karlsruhe, $76{\times}26$ mm) separated by a thin layer of 
silicone grease (GE Bayer Silicones Baysilone, medium viscosity) 
as spacer. The chambers were quasi-two-dimensional with a height 
ranging between 30 and 50 $\mu$m. The lateral sides were sealed 
by silicone grease after filling the cavity with bacterial 
suspension. Microscopy was performed with a Nikon Eclipse 
TE2000-s microscope and a Nikon 4x, N.A.\ 0.2 objective in 
the dark field mode. For each sample, sequences of images for 
2 min with 60 Hz frame rate were recorded using a Point Grey 
FL3-U3-88s2cc camera. The experiments were repeated with 
three different cultures for each strain.\\

\noindent\textbf{Trajectory selection and tumbling detection}\\
A triangular smoothing filter was applied to smooth the data. 
Circular trajectories in the vicinity of the surface, extremely 
short tracks, and abnormal ones belonging to the cells in late 
exponential or dividing phase were discarded. More than $2500$ 
cell trajectories were eventually selected for further analysis.

\begin{figure}[b]
\centerline{\includegraphics[width=0.48\textwidth]{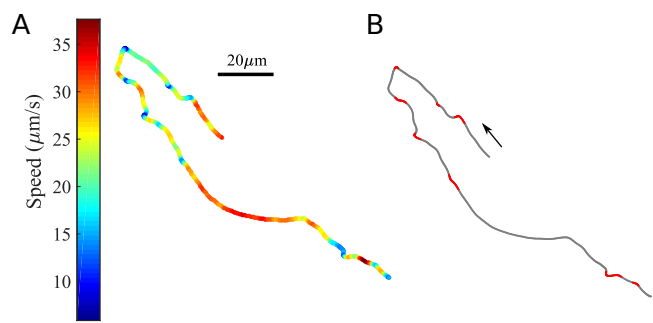}}
\caption{\textbf{Sample trajectory with detected tumbling events.} 
(\textbf{A}) The trajectory is color coded with respect to speed. 
(\textbf{B}) The detected tumbling events are indicated with red 
color. The arrow shows the direction of swimming.}
\label{Fig:5}
\end{figure}

The tumbling detection algorithm was based on identifying the 
dramatic changes in speed $v(t)$ and magnitude of angular velocity 
$\omega(t)$ of the bacterium \cite{Theves13,Masson12}. First, all 
local minima (maxima) of $v$ ($\omega$) over time were determined. 
Each minimum (maximum) is surrounded by two maxia (minima) 
located at $t_{1}$ and $t_{2}$. The depth $\Delta v$ of the 
minimum or the height $\Delta \omega$ of the maximum can be 
characterized as $\Delta v\!=\!\text{max}[v(t_{1}){-}v(t_\text{min}
),v(t_{2}){-}v(t_\text{min})]$ or $\Delta \omega\!=\!\text{max}
[\omega(t_\text{max}){-}\omega(t_{1}),\omega(t_\text{max}){-}
\omega(t_{2})]$, respectively. To identify a tumbling event 
two criteria were imposed: (i) If $\Delta v{/}v(t_\text{min})
{\geq}0.7$, there can be a tumbling phase around $t_\text{min}$. 
The possible tumbling period is limited to those times $t$ 
around $t_\text{min}$ where $v(t){-}v(t_\text{min}){\leq}0.2
\,\Delta v$. (ii) The local maximum of $\omega$ may indicate 
a tumbling event if the total directional change during the 
time interval $t_{2}{-}t_{1}$ exceeds $\sqrt{0.8(t_{2}{-}t_{1})}$. 
The corresponding tumbling period consists of those times $t$ 
around $t_\text{max}$ where $\omega(t_\text{max}){-}\omega(t){
\leq}\Delta \omega$. Imposing both conditions to identify a 
tumbling event ensures that the reduction of speed is 
accompanied by a sudden change in the direction of motion. 
See \cite{Theves13,Masson12} for more details. A sample 
trajectory with detected tumbling events is shown in Fig.\,\ref{Fig:5} 
(see also \emph{Suppl.\ Movie S1}).\\

\noindent\textbf{Analytical approach}\\
\noindent To describe the persistent motion of bacteria interrupted 
by stochastic tumbling periods, we developed an analytical framework 
for a random walker with two states of motility, as described by the 
set of master Eqs.~\ref{Eq:MasterEqs}. In the following, we briefly 
explain how arbitrary moments of displacement can be obtained by a 
Fourier-z-transform technique. Here, a two-dimensional motion is 
considered but extension to three dimensions is straightforward. 
The Fourier transform of $P_{t}^j(x,y|\gamma)$ is defined as
\begin{align}
P_{t}^j(\boldsymbol\omega|m) \equiv \int_{-\pi}^{\pi} \text{d}
\gamma\;e^{im\gamma} \int \!\! \text{d}y \int \!\! \text{d}x 
\,\, e^{i\boldsymbol\omega\cdot\boldsymbol r} P_{t}^j(x,y|\gamma),
\label{Eq:FourierDefinition_two_int}
\end{align}
with $j{\in}\{R,T\}$. An arbitrary moment of displacement 
$\langle x^{k_{1}} y^{k_{2}} \rangle^j(t)$ can be obtained as
\begin{align}
\langle x^{k_{1}} &y^{k_{2}} \rangle^j(t) \equiv  \int \!\! 
\text{d}\gamma\int \!\! \text{d}y \int \!\! \text{d}x \,\, 
x^{k_{1}} y^{k_{2}}  P_{t}^j(x,y|\gamma) \\ 
&= \left. \!\!\! (-i)^{k_{1}+k_{2}} \frac{\partial^{k_{1}+k_{2}} 
P_{t}^j(\omega_{x},\omega_{y}|m{=}0)}{\partial \omega_{x}^{k_{1}}
\partial \omega_{y}^{k_{2}}} \right|_{(\omega_{x},\omega_{y})=(0,0)}\!\!.
\end{align}
The master Eqs.~\ref{Eq:MasterEqs} after Fourier transformation read
\begin{equation}
\begin{aligned}
&P_{\!_{t{+}\Delta t}}^R(\omega, \alpha |m) = \\
&\sum_{k=-\infty}^{\infty} \! i^k\,e^{-ik\alpha}  
\,J_k(\omega \, v_{_\text{R}} \Delta t) \Big[f_{_{_{\text{T}{\!
\rightarrow\!}\text{R}}}} \, \mathcal{R}(m{+}k) \, P_{t}^T(
\omega,\alpha |m{+}k) \\
&+(1{-}f_{_{_{\text{R}{\!\rightarrow\!}\text{T}}}})\, p(m{+}k)
\,P_{t}^R(\omega,\alpha|m{+}k)\Big],
\label{Eq:MasterEqsFourier1}
\end{aligned}
\end{equation}
\begin{equation}
\begin{aligned}
&P_{\!_{t{+}\Delta t}}^T(\omega, \alpha |m) = \\
&\sum_{k=-\infty}^{\infty} \! i^k\,e^{-ik\alpha}  
\,J_k(\omega \, v_{_\text{T}} \Delta t) \Big[f_{_{_{\text{R}{\!
\rightarrow\!}\text{T}}}} \, P_{t}^R(\omega,\alpha |m{+}k) \\
&+(1{-}f_{_{_{\text{T}{\!\rightarrow\!}\text{R}}}})\, 
P_{t}^T(\omega,\alpha|m{+}k)\Big],
\label{Eq:MasterEqsFourier2}
\end{aligned}
\end{equation}
where we used the $k$th order Bessel's function
\begin{equation}
J_k(z) = \frac{1}{2\pi i^{k}} \int_{-\pi}^{\pi} \!\! \text{d}\gamma \,\, 
e^{iz\cos\gamma} e^{-ik\gamma},
\end{equation}
the Fourier transforms of the turning-angle distribution
\begin{equation}
\mathcal{R}(m)= \int_{-{\pi}}^{{\pi}} \!\!\! \text{d}\phi \,\, 
e^{i m \phi} R(\phi),
\end{equation}
and the distribution of the directional change along the 
run trajectory
\begin{equation}
p(m)= \int_{-{\pi}}^{{\pi}} \!\!\! \text{d}\theta \,\, 
e^{i m \theta} R(\theta).
\end{equation}
The Fourier transform of the probability $P_{t}^j(\omega,\alpha|m)$ 
can be expanded as a Taylor series
\begin{equation}
\begin{aligned}
P_{t}^j(\omega, \alpha|m) &= Q_{0,t}^j(\alpha|m) + i\,\omega 
\, v_{_\text{j}} \Delta t \,\, Q_{1,t}^j(\alpha|m) \\
&\hspace{4mm}- \frac{1}{2} \omega^2 \, v^2_{_\text{j}} (\Delta t)^2 
\,\, Q_{2,t}^j(\alpha|m)+ \cdot \cdot \cdot,
\end{aligned}
\end{equation}
and the moments of displacement can be read in terms of the Taylor 
expansion coefficients. For example,
\begin{align}
\langle x^2 \rangle^j(t) = \!  
v^2_{_\text{j}} (\Delta t)^2 \, Q_{2,t}^j(0|0).
\end{align}
Thus we expand both sides of the master equations \ref{Eq:MasterEqsFourier1} 
and \ref{Eq:MasterEqsFourier2} and collect all terms with the same 
power in $\omega$. As a result, coupled recursion relations for the 
Taylor expansion coefficients of terms with the same power in $\omega$ 
can be obtained. Next, the time indices on both sides of these equations 
can be equalized by means of $z$-transform, which enables us to obtain 
the moments of displacement in the $z$-space. For example, one obtains 
the following exact expression for the mean square displacement
\begin{equation}
\begin{aligned}
&\langle x^2 \rangle(z) {=} \!\!\!\sum_{t=0}^{\infty} z^{-t} \langle x^2 
\rangle(t) {=}\\ 
&(\Delta t)^2 \Big[ v^2_{_\text{R}} Q_{2}^R(z,0|0) + 
v^2_{_\text{T}} Q_{2}^T(z,0|0) \Big]=\\
&(\Delta t)^2 \Big[ (1{-}f_{_{_{\text{R}{\!\rightarrow\!}\text{T}}}}) Q_{0}^R(z,0|0) 
{+} f_{_{_{\text{T}{\!\rightarrow\!}\text{R}}}} Q_{0}^T(z,0|0) \Big] \times \\
& \Bigg[ \frac{z\left[z{-}(1{-}f_{_{_{\text{T}{\!\rightarrow\!}\text{R}}}})
\right]}{(z{-}1)G(z)} v^2_{_\text{R}} {+} \frac{z}{(z{-}1)G(z)} 
f_{_{_{\text{R}{\!\rightarrow\!}\text{T}}}} v_{_\text{R}} v_{_\text{T}} 
{-} \frac{1}{2(z{-}1)} v^2_{_\text{R}} \Bigg] \\
&{+} (\Delta t)^2 \Big[f_{_{_{\text{R}{\!\rightarrow\!}\text{T}}}} Q_{0}^R(z,0|0) 
{+} (1{-}f_{_{_{\text{T}{\!\rightarrow\!}\text{R}}}}) Q_{0}^T(z,0|0) \Big] \times \\
& \Bigg[\frac{z\left[z{-}(1{-}f_{_{_{\text{T}{\!\rightarrow\!}\text{R}}}})p
\right]}{(z{-}1)G(z)} v^2_{_\text{T}} {+} \frac{z}{(z{-}1)G(z)} 
f_{_{_{\text{T}{\!\rightarrow\!}\text{R}}}} \mathcal{R} \, v_{_\text{R}} v_{_\text{T}} 
{-} \frac{1}{2(z{-}1)} v^2_{_\text{T}} \Bigg],
\label{Eq:MSDz}
\end{aligned}
\end{equation}
where
\begin{equation}
\begin{aligned}
G(z)=&\left[z{-}(1{-}f_{_{_{\text{T}{\!\rightarrow\!}\text{R}}}})\right]
\left[z{-}(1{-}f_{_{_{\text{R}{\!\rightarrow\!}\text{T}}}})p\right]-
f_{_{_{\text{R}{\!\rightarrow\!}\text{T}}}} f_{_{_{\text{T}{\!\rightarrow\!}
\text{R}}}} \mathcal{R}.
\end{aligned}
\end{equation}
By inverse $z$-transforming the moments of displacement in the $z$-space, 
the moments can be obtained as a function of time.\\ 

\noindent\textcolor{red}{SUPPLEMENTARY MATERIALS}\\
\noindent movie S1. A sample bacterial trajectory.

\smallskip\smallskip\footnotesize{
\noindent\textbf{Acknowledgments:} We thank Daniel B.\ Kearns 
for providing us with strain DK1693, and Z.\ Sadjadi for fruitful 
discussions. \textbf{Funding:} M.\,R.\,S.\ acknowledges support 
by DFG within SFB 1027 (A7). \textbf{Author contributions:} 
C.\,W. and G.\,B. designed the research. F.\,A. prepared the strains 
of the bacterium Bacillus subtilis. C.\,W., T.\,J., and J.\,N. designed 
and built the experimental setup. J.\,N. carried out the experiments 
and analyzed the results. M.\,R.\,S. developed the analytical model 
and performed simulations. All authors contributed to the interpretation 
of the results. M.\,R.\,S. drafted the manuscript. \textbf{Competing 
interests:} The authors declare that they have no competing interests. 
\textbf{Data and materials availability:} All data needed to evaluate 
the conclusions in the paper are present in the paper. Additional data 
related to this paper may be requested from the authors.}

\end{document}